\documentclass[amsmath, amssymb,10pt,aps,prb,twocolumn,notitlepage,showpacs,superscriptaddress]{revtex4-1}
\usepackage{graphicx}
\usepackage{calc}
\usepackage{braket}
\usepackage{ulem}
\usepackage{slashed}
\usepackage{wasysym}
\usepackage{siunitx}
\usepackage{dsfont}
\usepackage{amsthm,amsmath,amsfonts,amssymb,verbatim,color}
\usepackage{graphicx}
\usepackage{subfigure}
\usepackage{bm}
\usepackage{epsfig,slashed}
\usepackage[T1]{fontenc}
\usepackage[colorlinks=true,citecolor=blue,linkcolor=blue,urlcolor=blue]{hyperref}
\usepackage[version=4]{mhchem}
\begin{document}
\title{Giant intrinsic  anomalous terahertz Faraday rotation in the magnetic Weyl semimetal Co$_2$MnGa at room temperature }

\author{Xingyue Han}
\affiliation{Department of Physics and Astronomy, University of Pennsylvania, Philadelphia, Pennsylvania 19104, USA}
\author{Anastasios Markou  }
\affiliation{Max-Planck-Institut fur Chemische Physik fester Stoffe, 01187 Dresden, Germany}
\author{Jonathan Stensberg}
\affiliation{Department of Physics and Astronomy, University of Pennsylvania, Philadelphia, Pennsylvania 19104, USA}
\author{Yan Sun  }
\affiliation{Max-Planck-Institut fur Chemische Physik fester Stoffe, 01187 Dresden, Germany}
\author{Claudia Felser}
\affiliation{Max-Planck-Institut fur Chemische Physik fester Stoffe, 01187 Dresden, Germany}
\author{Liang Wu}
\email{liangwu@sas.upenn.edu}
\affiliation{Department of Physics and Astronomy, University of Pennsylvania, Philadelphia, Pennsylvania 19104, USA}

\date{\today}

\begin{abstract}

We report measurement of terahertz anomalous Hall conductivity and Faraday rotation in the magnetic Weyl semimetal Co$_2$MnGa thin films as a function of the magnetic field, temperature and thickness, using time-domain terahertz spectroscopy. The terahertz  conductivity shows a thickness-independent anomalous   Hall conductivity of around 600 $\Omega^{-1}\cdot cm^{-1}$ at room temperature, and it is also frequency-independent from 0.2-1.5 THz. The magnitude of the longitudinal and Hall conductivities, the weak spin-orbit coupling, and  the very close positioning of Weyl points to the chemical potential all satisfy the criteria for intrinsic anomalous Hall conductivity. First-principle calculation also supports the frequency-independent intrinsic anomalous Hall conductivity at low frequency. We also find a thickness-independent Faraday rotation of 59 ($\pm6$) mrad  at room temperature, which comes from the intrinsic Berry curvature contribution. In the thinnest  20 nm sample, the Faraday rotation divided by the sample thickness reaches around 3 mrad/nm due to Berry curvature, and is the largest reported at room temperature. The giant Verdet constant of the order of 10 $^{6}$ rad m $^{-1}$ T $^{-1}$ at room temperature and the large Hall angle around 8.5 $\%$ from 0.2-1.5 THz indicates that Co$_2$MnGa  is promising for THz spintronics  at room temperature.

\end{abstract}

\maketitle

\textbf{Introduction} 

The last decade witnessed an explosion of research on topological states of matter characterized by the topological properties of the bulk wave-functions\cite{moore2010birth,HasanKaneRMP10, QiZhangRMP11}. Topological insulators are robust to adiabatic perturbation as long as the bulk gap is not closed\cite{HasanKaneRMP10, QiZhangRMP11}.  Weyl semimetals are newly discovered topological states of matter without a bulk gap and with open Fermi surface arcs when either time-reversal or inversion symmetry is broken\cite{WanPRB2011, BurkovPRL2011, XuPRL2011, WengPRX2015, HuangNatComm2015, XuScience2015, LvPRX2015, YangNatPhys2015}. These materials have accidental band touching at pairs of points with different chirality in the momentum space. Near these points, the quasi-particles (low-energy excitations) can be described by Weyl equations first proposed by Hermann Weyl in 1929\cite{weyl1929elektron}. As a result, these touching points in the band structure are called Weyl points, and the quasi-particles near them resemble Weyl fermions. The bulk wave functions in Weyl semimetals acquire a Berry phase as they move around the Weyl point because each Weyl node behaves like a fictional magnetic field known as Berry curvature. These Weyl points (monopoles in k space) are also topologically protected because translation-invariant perturbations are identical to moving Weyl points in the momentum space, unless they meet in the zone boundary and annihilate with each other.  

Weyl semimetals host many exotic phenomena including interesting temperature and frequency dependence in optical conductivity\cite{HosurPRL2012,xu2020PNAS,ni2020NPJ}, novel quantum oscillations related with Fermi arcs\cite{PotterNatComm2014}, giant second harmonic generation\cite{wu2017NatPhy,patankarPRB2018}, as well as a chiral magnetic effect and intrinsic anomalous Hall effect\cite{BurkovJPCM2015}.  Experimental signatures of Weyl semi-metals are still far behind the advance of various theoretical proposals.   Thus far, the widely experimentally studied Weyl semi-metal materials are non-magnetic, but with broken inversion symmetry\cite{XuScience2015, LvPRX2015, YangNatPhys2015, wuNatPhys2017}. Fermi arcs in magnetic Co$_3$Sn$_2$S$_2$ and Co$_2$MnGa  were identified recently \cite{belopolskiSci2019,moraliSci2019}, which established direct evidence for the magnetic Weyl semimetals. Among them, Co$_2$MnGa is particularly interesting as it is a room temperature ferromagnetic with a large anomalous Hall effect and a high curie temperature at $T_C$=690 K \cite{mannaPRX2018,markouPRB2019}. Bulk Co$_2$MnGa  is a Heusler compound, which has a cubic face-centered structure with space group $Fm\bar3m$ (No. 225).  Large  anomalous Nernst effects  are also observed in both bulk and thin films\cite{sakaiNatPhy2018,reichlovaAPL2018,parkPRB2020}. 

The anomalous Hall effect (AHE)  has been studied extensively in conventional ferromagnetic materials \cite{nagaosaRMP2010,leeSci2004}.  The intrinsic AHE is a scattering-independent process first proposed by Karplus and Luttinger \cite{karplusPR1954}, which depends only on the topological band structure with the contribution of the Berry curvature in the momentum space \cite{nagaosaRMP2010,xiaoRMP2010,armitageRMP2018}. The AHE can be greatly enhanced when Fermi energy is close to band (anti-)crossings such as the Weyl points \cite{fangSci2003,yaoPRL2004,kublerEPL2016}. Due to its exotic band structure, magnetic Weyl semimetals (WSMs) such as Co$_2$MnGa are a good platform to study the intrinsic AHE  \cite{yanARCMP2017,hosurCRP2013}.  When the chemical potential is close to the Weyl points, the nonzero net Berry curvature effect is dominating, and  a large intrinsic anomalous Hall effect can be calculated by the Kubo formula and compared with experiments \cite{nagaosaRMP2010}. 

Another advantage of Co$_2$MnGa is that thin films, which are ideal for device applications, can be fabricated by sputtering. The samples studied in this work are grown by sputtering on MgO substrates and capped with 3 nm Al, the growth details of which were the subject of previous study\cite{markouPRB2019}.  At room temperature, transport measurement on thin films show a large anomalous Hall conductivity (AHC) of 814 $\Omega^{-1}\cdot cm^{-1}$ and a large anomalous Hall angle (AHA) of 10.5 \%\cite{markouPRB2019}.  However, the spectrum of the anomalous Hall conductivity, $\sigma_{xy}(\omega)$, and Hall angle, $\theta_H(\omega)$, at terahertz frequency remain missing. Future high-speed spintronic devices will crucially rely on such AHE phenomena at terahertz (THz) frequencies.  Current DC AHE transport measurements significantly lag behind other interests of information carriers such as electrons in field-effect transistors featuring cut-off frequencies of around 1 THz \cite{delnature2011}. Therefore, a proper understanding of the thin-film properties at THz frequencies is required. Another intriguing aspect is the possibly giant magneto-optical Faraday effect, i.e. the rotation of the light polarization of the transmitted light for the magnetic media, resulting from the enhanced Hall conductivity on the transitions near the Weyl point. The magnitude of the Faraday rotation is proportional to the terahertz Hall conductivity. This novel optical property inherent to the magnetic WSM remains to be explored.

In our work, we study the THz AHC, AHA and Faraday rotation spectra of Co$_2$MnGa  thin films using magneto-optical methods. We employ time-domain terahertz spectroscopy (TDTS) to measure optical longitudinal conductivity and complex Faraday rotation at 0.2-1.5 THz. Giant THz AHC and AHA at RT around 600 $\Omega^{-1}\cdot cm^{-1}$ and 8.5 \% are observed. AHE remains operative from DC up to 1.5 THz with a flat frequency response in thin films. The optical Faraday rotation and AHC both show a thickness-independent behavior at RT. One of the samples exhibits a 2.5 mrad/nm Faraday rotation divided by thickness, which is the largest value reported at RT. Using first-principle calculation, we show that the AHC and Faraday rotation are dominating from the intrinsic contribution.
\\

\begin{figure}
\centering
\includegraphics[width=0.5\textwidth]{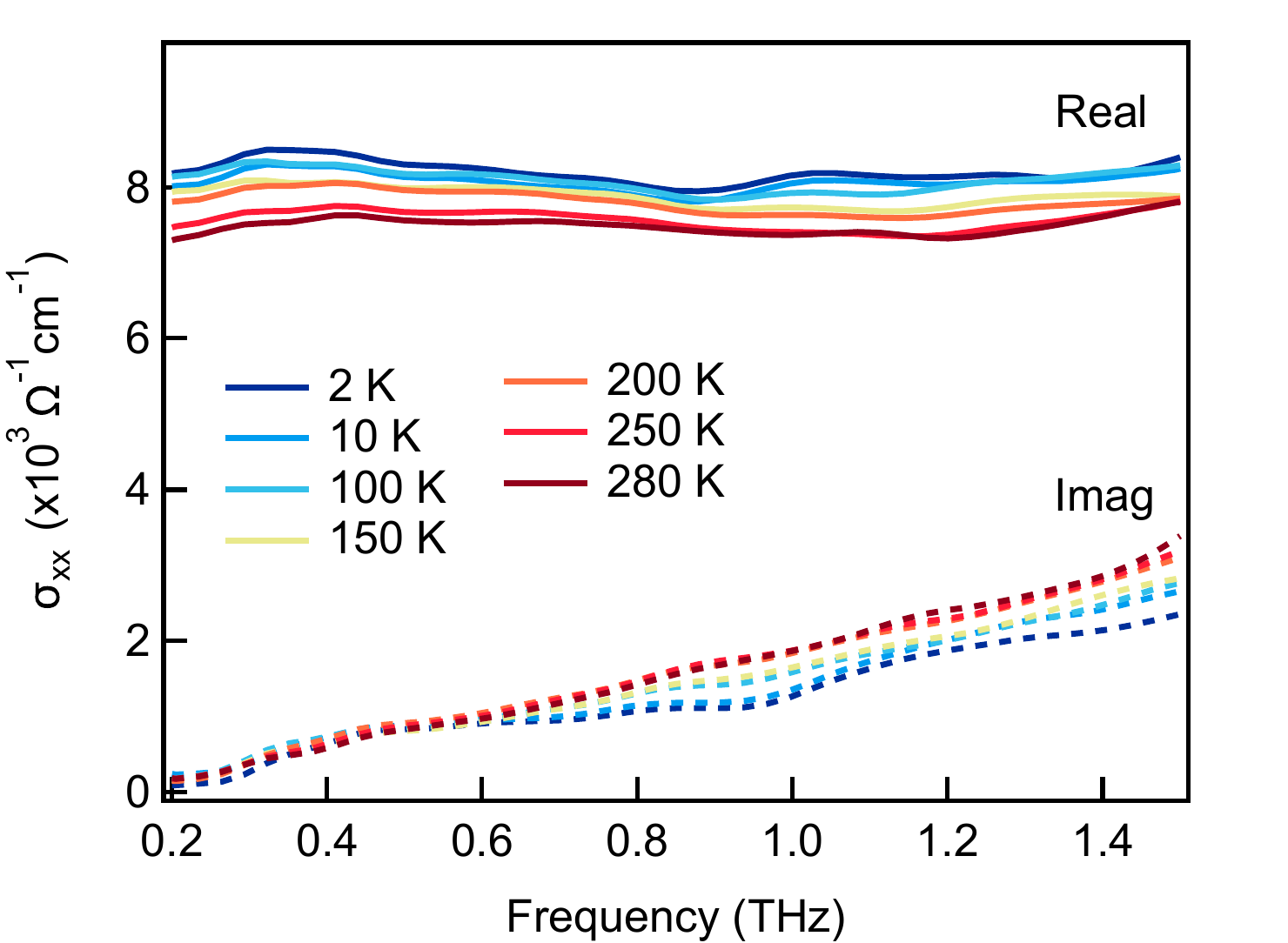}
\caption{ Longitudinal conductivity spectra $\sigma_{xx}(\omega)$ of a 40 nm Co$_2$MnGa thin film  from 2 K to 280 K. The solid lines are the real part. The dashed lines are the imaginary part.}
\label{Fig1}
\end{figure}

\textbf{Results and Discussion}

\begin{figure*}
\centering
\includegraphics[width=\textwidth]{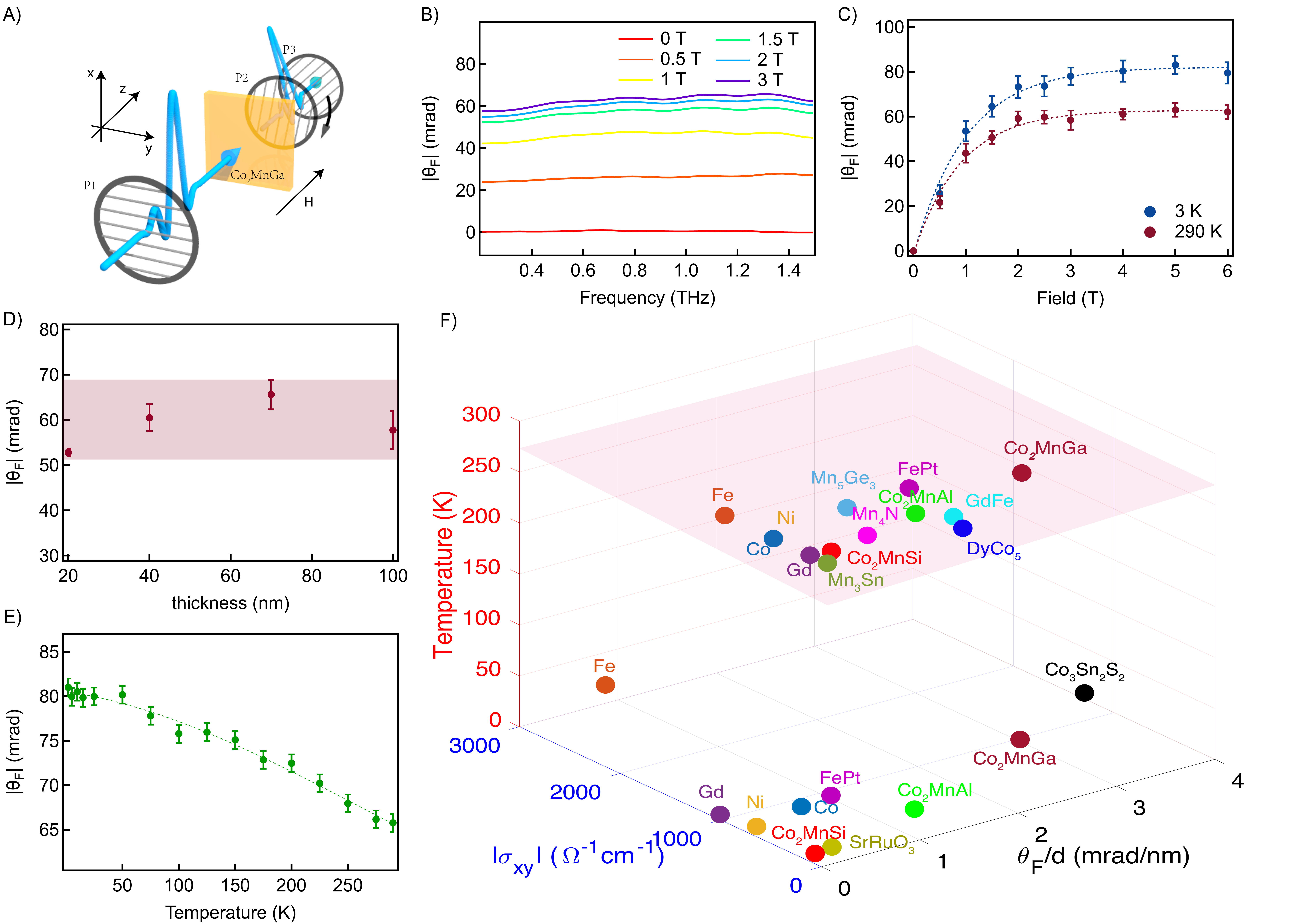}
\caption{ A) A sketch of the experimental geometry of the Faraday rotation measurement. B) Field dependent Faraday angle spectra of a 40 nm Co$_2$MnGa sample  at 290 K. C) Field dependence of the average Faraday angle over the frequency range of 0.2-1.5 THz of the 40 nm Co$_2$MnGa sample. D) Thickness independent Faraday angle under 2 T at 290 K. E) Temperature dependence of Faraday angle of the 40 nm Co$_2$MnGa  under 2.5 T. F) Thickness-normalized Faraday rotation and anomalous Hall conductivity for Co$_2$MnGa and several magnetic thin films. The values of \ce{Co2MnGa}, \ce{Co3Sn2S2}\cite{okamuraNatComm2020}, \ce{Mn3Sn}\cite{matsudaNatComm2020,Khadka2020SciAdv}, and \ce{SrRuO3}\cite{shimano2011EPL} are measured by THz measurement. The values of GdFe\cite{seifert2021AdvMat} and \ce{DyCo5}\cite{seifert2021AdvMat} are estimated from THz measurement. The values of Fe\cite{miyasatoPRL2007}, Co\cite{miyasatoPRL2007}, Ni\cite{miyasatoPRL2007}, Gd\cite{miyasatoPRL2007}, FePt\cite{he2012PRL}, \ce{Co2MnAl}\cite{sakuraba2020PRB}, \ce{Co2MnSi}\cite{sakuraba2020PRB}, \ce{Mn5Ge3}\cite{zeng2006PRL} and \ce{Mn4N}\cite{ shen2014APL} are estimated from transport measurement. }
\label{Fig2}
\end{figure*}

In our TDTS, a laser beam from a fiber laser (center wavelength 780nm, duration 82 fs) is split by a beam splitter into pump and probe beams. On the pump side, a photoconductive antenna with 34 V bias voltage and 1.8 kHz modulation frequency is used to generate THz pulses. Four off-axis parabolic mirrors (OAPs) are placed in a confocal geometry to collect and focus the THz beam. Thin film sample is put on the focus of the second and third OAP. On the probe side, the time delayed probe beam hits the detector antenna to detect the transmitted THz electric field using optical sampling method. The entire THz path is enclosed in a purge box with relative humidity <3\% to avoid water absorption. We present the zero-field terahertz conductivity on these films first. Utilizing TDTS, we obtain the complex longitudinal conductivity spectra, $\sigma_{xx}(\omega)$, as we measure both the amplitude and the phase of the transmission without using Kramers–Kronig transformation \cite{WuNatPhys13,WuPRL2015}. In the thin-film approximation, the longitudinal conductivity can be extracted from the following relation:
\begin{equation}
T_{xx}(\omega)=\dfrac{n+1}{n+1+d\sigma_{xx}(\omega)Z_0}exp(i\dfrac{\omega}{c}\Delta L(n-1)).
\end{equation}
Here $T_{xx}$ is the transmission coefficient of the thin film,  $n$ is the refractive index of the substrate, $d$ is the thickness of the film, $Z_0$ is the impedance of free space, $\Delta L$ is the thickness mismatch between the sample and the bare substrate. Fig. \ref{Fig1} shows the real and imaginary part of $\sigma_{xx}(\omega)$ of 40 nm Co$_2$MnGa  from 2 K to 280 K. The temperature dependence is quite weak as the the real conductivity changes only by 8 $\%$ from 2 K to 280 K. The results follow the Drude model, $\sigma_{xx}(\omega)=\sigma_0/(1-i\omega\tau)-i\epsilon_0(\epsilon_\infty-1)\omega$, where $\sigma_0$ is the DC conductivity,  $\tau$ is the scattering time, $\epsilon_0$ is the vacuum permittivity, and $\epsilon_\infty$ is a constant that describes lattice polarizability at high frequency. The flat spectra in the real part (solid lines) and the small linear spectra of Im $\sigma_{xx}(\omega)$ (dashed lines) both indicates a short $\tau$. Fitting shows $1/\tau$ is around 3-5 THz in the whole temperature range. The magnitude of the real part of the longitudinal conductivity falls into one of the criteria for intrinsic AHE \cite{miyasatoPRL2007,onodaPRL2006} as discussed below.

Besides the intrinsic contribution to AHE, there are two other kinds of extrinsic contributions: skew scattering and side jump \cite{nagaosaRMP2010}. In ferromagnets, the combination of the total anomalous Hall and longitudinal conductivity show a crossover behavior in three regions according to the magnitude of longitudinal conductivity\cite{miyasatoPRL2007,onodaPRL2006}: dirty regime ($\sigma_{xx}<10^4$ $\Omega^{-1}\cdot cm^{-1}$), intermediate regime ($\sigma_{xx}=10^4-10^6$ $\Omega^{-1}\cdot cm^{-1}$), and extreme conducting regime ($\sigma_{xx}>10^6$  $\Omega^{-1}\cdot cm^{-1})$. The intrinsic contribution always dominates in the intermediate regime. The magnitude of the real part of the longitudinal conductivity in Co$_2$MnGa falls at the lower boundary in the intermediate regime. Often, to separate the three contributions, the scaling relation between longitudinal conductivity and anomalous Hall conductivity is used. The contrition of the skew scattering to the AHC is proportional to the square of longitudinal conductivity $\sigma_{xy}^{A-skew}\propto\sigma_{xx}^2$, while the intrinsic and side-jump contributions are both independent of the longitudinal conductivity. Previous DC transport on Co$_2$MnGa bulk crystal used the scaling relation and revealed that the intrinsic AHC is around $10^3$ $\Omega^{-1}\cdot cm^{-1}$ \cite{mannaPRX2018,belopolskiSci2019}.

Nevertheless, it is very difficult to separate the intrinsic contribution and side jump in DC transport \cite{nagaosaRMP2010}. Interestingly, Co$_2$MnGa also satisfies the other two criteria that favors intrinsic contribution \cite{onodaPRL2006}. 1) The AHC is on the order of 1000 $\Omega^{-1}$$\cdot$$cm^{-1}$. The intrinsic AHC value in the thin film and bulk crystal was reported to be 1138 $\Omega^{-1}$$\cdot$$cm^{-1}$\cite{markouPRB2019} and 1164 $\Omega^{-1}$$\cdot$$cm^{-1}$\cite{mannaPRX2018} respectively. 2) The anticrossing point (the Weyl point) is only around 80 meV above the chemical potential\cite{belopolskiSci2019}. Because of these three criteria, one can perform first-principles calculations to reliably predict the frequency-dependent intrinsic contribution and compare it with experiments.

With a set of freestanding wire-grid polarizers as shown in Fig. \ref{Fig2}A, our TDTS can resolve the polarization state of the THz signal and measure the frequency dependent AHC. Three THz wire-grid polarizers (P1,P2,P3, extinction ratio>2000 at 1 THz) are used to measure the Faraday angle. P1 aligns the incident polarization vertically. P2 is mounted on a rotation stage (not shown in the figure) to selectively pass the vertical electric field ($E_x$) or horizontal electric field ($E_y$). P3 is fixed at 45$^\circ$ so that $E_x$ and $E_y$ have the same response at the detector. The polarization change before and after the Co$_2$MnGa  film is the Faraday rotation $\theta_F(\omega)=\dfrac{E_y(\omega)}{E_x(\omega)}$. These measurements were performed under an out-of-plane magnetic field up to 7 T. To exclude the nonmagnetic effects such as birefringence from the windows, we apply $\pm$B to get a symmetrized Faraday angle spectra $\theta_F(\omega,H)=[\theta_F^{meas}(\omega,H)-\theta_F^{meas}(\omega,-H)]/2$. Fig. \ref{Fig2}B shows the Faraday angle $\theta_F (\omega)$ of the 40 nm sample at 290 K. The rotation has a weak dependence on the frequency. The field dependence is similar to the known magnetization curve with a saturation field $H_s\approx1.5$ T\cite{markouPRB2019} as shown in Fig. \ref{Fig2}C. Note that usually the Faraday rotation is proportional to the thickness and characterized by the Verdet constant \cite{meyers1987encyclopedia}.  The large $\theta_F$ around 60 mrad at room temperature is nearly thickness independent as in Fig. \ref{Fig2}D, which mainly contribute from the terahertz anomalous Hall effect. $\theta_F$ increases monotonically  and reaches 80 mrad as the temperature cools down to 2K  (see Fig. \ref{Fig2}E).  Below 1 Tesla, the Verdet constant of the 40 nm sample reaches  10 $^{6}$ $rad$ $m ^{-1} T ^{-1}$, which is of similar size of the giant magneto-optical effect in topological insulators HgTe \cite{shuvaevPRL2011} and Bi$_2$Se$_3$ \cite{ValdesAguilarPRL12, WuPRL2015} at low temperature. 

In Fig. \ref{Fig2}F, we show the normalized Faraday rotation by thickness of a few magnetic materials at room temperature and low temperature. Co$_2$MnGa has the largest room temperature value of 3 mrad/nm to our best knowledge. It is 2-3 orders of magnitude larger than the typical transition metals Fe, Co, Ni, and Gd\cite{miyasatoPRL2007}. It is 15-30 times larger than the value in the Weyl antiferromagnet candidate Mn$_3$Sn \cite{matsudaNatComm2020,Khadka2020SciAdv} and the ferromagnet Co-based Heusler compound Co$_2$MnSi \cite{sakuraba2020PRB}. The value in \ce{Co2MnGa} is twice larger than another half metallic ferromagnet, \ce{Co2MnAl} \cite{sakuraba2020PRB}, which is known to exhibit similar AHE from bulk crystal with \ce{Co2MnGa}\cite{li2020NatComm}. Even if we make the comparison at low temperature,  3 mrad/nm is similar to the magnetic Weyl semimetal Co$_3$Sn$_2$S$_2$\cite{okamuraNatComm2020}, and larger than any other materials (see Fig.\ref{Fig2}F). The realization of the giant Faraday rotation at room temperature overcomes the cryogenic temperature limit, which makes it promising in spintronics applications.  A recent work measures the broadband THz anomalous Hall conductivity  in the ferrimagnets DyCo$_5$ and GdFe \cite{seifert2021AdvMat}, and we compute the Faraday rotation by using the reported transverse and longitudinal conductivity. The small anomalous Hall conductivity in DyCo$_5$ probably  indicates an extrinsic origin from the side jump or skewing scattering. GdFe is interesting as it also shows both large  anomalous Hall conductivity and Faraday rotation, but whether the origin of both are intrinsic needs further investigation. Note that the GdFe \cite{seifert2021AdvMat} samples studied previously were amorphous. Therefore the Berry curvature from the Bloch wave function is ill-defined, and  it is not straight-forward to study the intrinsic effect in GdFe \cite{seifert2021AdvMat}. As discussed below, we will show that the large  anomalous Hall conductivity and Faraday rotation have an intrinsic orgin in Co$_2$MnGa.   

\begin{figure*}
\includegraphics[width=\textwidth]{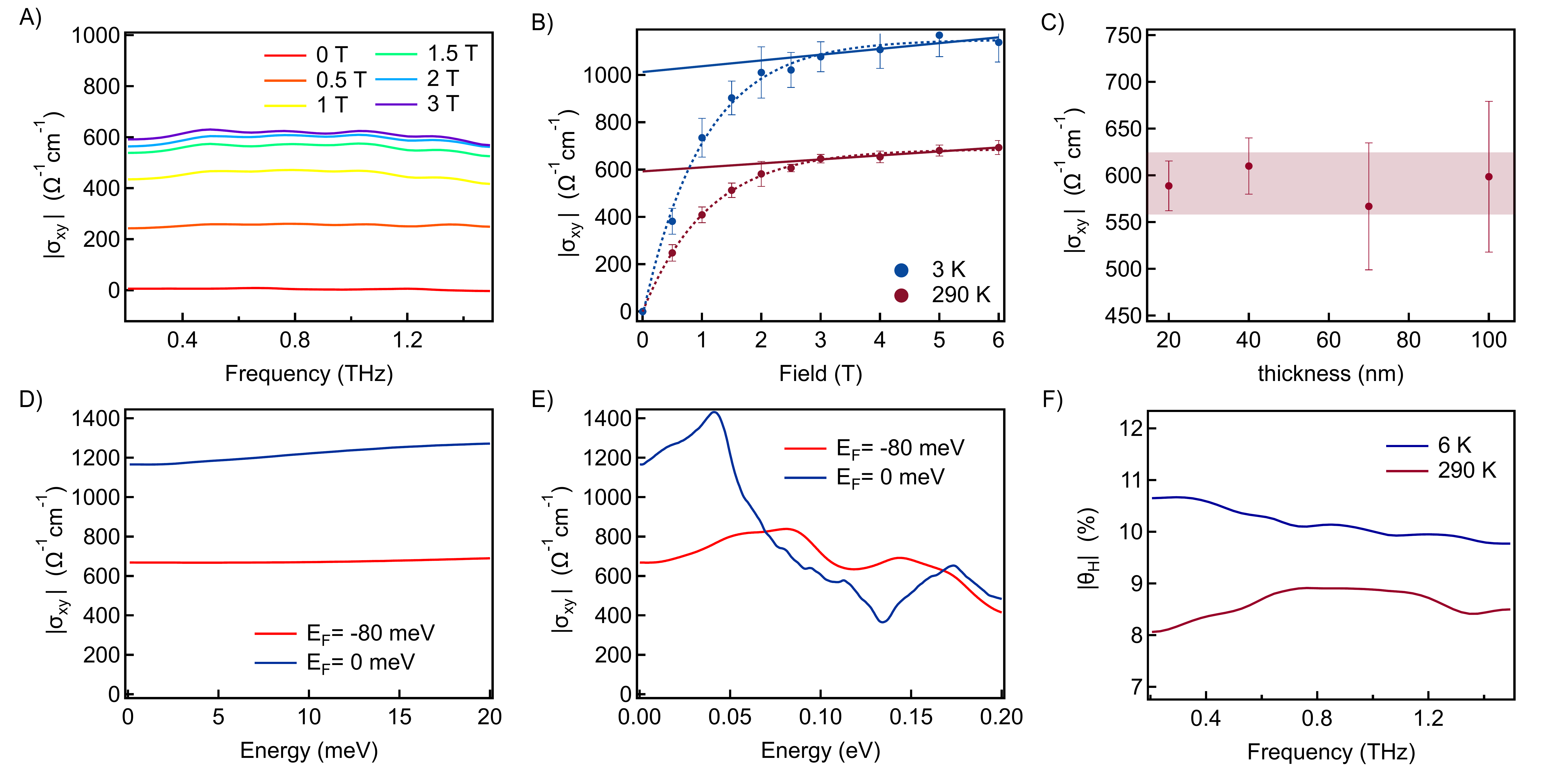}
\caption{ A) Field dependent Hall conductivity spectra of the 40 nm Co$_2$MnGa  at 290 K. B) Field dependence of the average Hall conductivity over frequenncy range of 0.2-1.5 THz of the 40nm Co$_2$MnGa  at 3 K and 290 K. The intercepts of dashed lines on the vertical axis are the anomalous Hall conductivity $\sigma_{xy}^A$. C) Thickness independent Hall conductivity $\sigma_{xy}$ under 2 T at 290 K. D,E) First principle calculation of RT Hall conductivity with $E_F$=-80 meV and 0 meV. We define the Weyl point position as 0 meV. F)Hall angle spectra of the 40 nm Co$_2$MnGa  at 6 K and 290 K}.
\label{Fig3}
\end{figure*}

The terahertz Hall conductivity spectra $\sigma_{xy} (\omega)$ can be extracted from the Faraday rotation by the relation 
\begin{equation}
\sigma_{xy}(\omega)=\theta_F(\sigma_{xx}+\dfrac{n+1}{dZ_0})
\end{equation} 
where $n$ is the refractive index of the substrate, $Z_o$ is the vacuum impedance and $d$ is the thickness of the film. FIG. \ref{Fig3}A shows the terahertz Hall conductivity $\sigma_{xy} (\omega)$ of 40 nm Co$_2$MnGa. Consistent with the flat spectra of longitudinal conductivity and Faraday rotation, it is also independent of frequency. The mean value between 0.2 and 1.5 THz under each field is plotted in FIG. \ref{Fig3}B, again scaling with the magnetization\cite{markouPRB2019}. According to the Hall conductivity formula:
\begin{equation}
\sigma_{xy}=R_0H+\sigma_{xy}^A
\end{equation}
The total Hall conductivity $\sigma_{xy} (\omega)$ consists of the ordinary Hall effect, $R_0H$, where $R_0$ is the ordinary Hall coefficient, and the anomalous Hall effect, $\sigma_{xy}^A$. The ordinary Hall conductivity scales with the magnetic field. The anomalous term can be obtained by extrapolating the high field data to zero field (FIG. \ref{Fig3}B intersection). The 40 nm Co$_2$MnGa  shows a giant value of $|\sigma_{xy}^A|=930$ $\Omega^{-1}\cdot cm^{-1}$ at 2 K, and $|\sigma_{xy}^A|=590$ $\Omega^{-1}\cdot cm^{-1}$ at 290 K. These results are consistent with the DC transport values on the 40 nm Co$_2$MnGa \cite{markouPRB2019}. Similar to RT Faraday rotation, the RT Hall effect is also thickness independent, with the Hall conductivity value  around 600 $\Omega^{-1}\cdot cm^{-1}$ as shown in FIG. \ref{Fig3}C. This can be qualitatively analysed from Eq.(2). As the sample thickness increases, the first term, $\sigma_{xx}$, slightly increases, but the second term $\frac{n+1}{dZ_0}$ decreases slightly as well. The changes in the two terms tend to cancel each other, resulting in a thickness independence behavior in the 20 to 100 nm range. However, the main reason for the thickness-independent Faraday rotation is due to the thickness-independent intrinsic anomalous Hall conductivity as discussed below. We would like to point out that  substrates from different vendors and different capping could lead to a change of 10$\%$ of AHC at room temperature. Note that the experiment value $|\sigma_{xy}^{exp}|\sim600$ $\Omega^{-1}\cdot cm^{-1}$ is very close to the theoretical prediction of intrinsic Hall effect in the order of $|\sigma_{xy}^{int}|\sim e^2/ha\sim670$ $\Omega^{-1}\cdot cm^{-1}$ for Co$_2$MnGa \cite{miyasatoPRL2007}, where $a$ is the lattice constant. This indicates that the intrinsic contribution (the Berry curvature) is the leading effect. 

The large scattering rate in Co$_2$MnGa makes it difficult to separate the intrinsic and side jump contributions via the scaling of $\sigma_{xy}$ versus  $\sigma_{xx}^2$ at terahertz frequency, as the measured THz frequency range is comparable to the scattering rate \cite{nagaosaRMP2010}.   Nevertheless, as we discussed above, because Co$_2$MnGa satisfies the three criteria for intrinsic anomalous Hall conductivity in terms of longitudinal conductivity, Hall conductivity, and chemical potential, it is quite accurate to use density functional theory (DFT) to calculate and identify the intrinsic contribution. We calculate the electronic band structure based on DFT by employing the full-potential local-orbital code (FPLO) with localized atomic basis~\cite{Koepernik1999}. The exchange and correlation energies were considered in the generalized gradient approximation (GGA) level following the Perdew–Burke–Ernzerhof parametrization scheme\cite{perdew1996}. Following experimental results, we set a ferromagnetic structure with magnetic moment along $z$ direction\cite{mannaPRX2018}. We projected the Bloch wavefunction into high symmetric atomic-orbital-like Wannier functions~\cite{koepernik2021} and constructed the tight-binding model Hamiltonian by the Wannier function overlap. Based on the tight-binding model Hamiltonian, the DC anomalous Hall conductivity  and THz conductivity were computed by following the Kubo formula approach in linear response approximation and clean limit~\cite{kubo1957,greenwood1958,xiaoRMP2010,nagaosa2010}, with AHC    
\begin{equation}
      \begin{aligned}
 \sigma_{xy}^{DC}(E_{F})=2e^{2}\hbar\int_{BZ}\frac{d^{3}k}{(2\pi)^{3}}\underset{E_{n}(k)\leq E_{F}}{\sum}f_{n}(k) \\ Im\underset{m\neq n}{\sum}\frac{<u_{n}(k)|\hat{v}_{x}|u_{m}(k)><u_{m}(k)|\hat{v}_{y}|u_{n}(k)>}{(E_{m}(k)-E_{n}(k))^{2}+\eta^{2}}
	  \end{aligned}
      \label{AHC}
      \end{equation}
and terahertz conductivity
     \begin{equation}
        \begin{aligned}
  \sigma_{xy}(\hbar\omega)=ie^{2}\hbar\int_{BZ}\frac{d^{3}k}{(2\pi)^{3}}\underset{m\neq n}{\sum}\frac{f_{m}(k)-f_{n}(k)}{E_{m}(k)-E_{n}(k)}\\\frac{<u_{n}(k)|\hat{v}_{x}|u_{m}(k)><u_{m}(k)|\hat{v}_{y}|u_{n}(k)>}{E_{m}(k)-E_{n}(k)-(\hbar\omega+i\eta)}
	  \end{aligned}
        \label{OC}
        \end{equation}
where $f_n(k)$ is the Fermi-Dirac distribution, $E_n(k)$ is the eigenvalue of $n$th band with eigenstate $|u_n(k)>$, 
$\hat{v}_{x}(k)=\frac{1}{\hbar}\frac{\partial\hat{H}(k)}{\partial k_{x}}$ is the velocity operator, $\hbar\omega$ is the transition energy, and $\eta$ is a smearing parameter to avoid numerical divergence. (Here we set $\eta$=0.1 meV.) We used a dense $k$-grid of $240^3$ for the numerical integration. As shown in FIG. \ref{Fig3}D, when the Weyl points are around 80 meV above the chemical potential, it matches the anomalous THz Hall conductivity, which also agrees with previous transport and ARPES studies \cite{markouPRB2019, belopolskiSci2019}. FIG. \ref{Fig3}D shows the intrinsic anomalous THz Hall conductivity over a larger frequency range with resonant features associated with Berry curvature contribution, which we hope future experiments could explore.

 We also measure the Hall angle at THz frequency, $\theta_H=\dfrac{\sigma_{xy}(\omega)}{\sigma_{xx}(\omega)}$, as it will be useful for field-effect transistors around 1 THz \cite{delnature2011}. A large Hall angle $\sim8.5\%$ is observed at 290 K, as shown in FIG. \ref{Fig3}F. It increases as temperature decreases and reaches a maximum with 10 \% at 2 K.  Co$_2$MnGa  is one of few materials that exhibits large AHE and large Hall angle at the same time. Since AHE and spin Hall effect  are generated by the same mechanisms, the large AHE in Co$_2$MnGa  guarantees a large spin Hall effect, which was reported recently\cite{leiva2021PRB}. Looking forward, we believe that our observation of large THz anomalous Hall conductivity, Faraday rotation, and Hall angle from the intrinsic contribution at room temperature will be critical to use Co$_2$MnGa for future topological spintronics at the THz frequencies.  \\

X.H., J.S. and L.W. acknowledge the support for this project from the ARO under the Grant W911NF1910342.  The acquisition of the cryostat is supported by the ARO under the Grant W911NF2020166. X.H. and J.S. are also partially supported by the Gordon and Betty Moore Foundation’s EPiQS Initiative, Grant GBMF9212 to L.W.  The acquisition of the laser for the THz system is  support from a seed grant at NSF supported University of Pennsylvania Materials Research Science and Engineering Center (MRSEC)(DMR-1720530). L.W. acknowledges partial summer support from the NSF EAGER grant  (DMR-2132591).

\bibliography{mainv2_clean.bib}

\end{document}